\documentstyle[12pt,epsfig]{article}
\def\IJMP #1 #2 #3 {{\it Int.\ J.\ Mod.\ Phys.}\ {\bf #1}\ (#2) #3}
\def\MPL #1 #2 #3 {{\it Mod.\ Phys.\ Lett.}\ {\bf #1}\ (#2) #3}
\def\NC #1 #2 #3 {{\it Nuovo Cim.}\ {\bf #1} (#2) #3}
\def\NP #1 #2 #3 {{\it Nucl.\ Phys.}\ {\bf #1}\ (#2) #3}
\def\PL #1 #2 #3 {{\it Phys.\ Lett.}\ {\bf #1}\ (#2) #3}
\def\PR #1 #2 #3 {{\it Phys.\ Rev.}\ {\bf #1}\ (#2) #3}
\def\PP #1 #2 #3 {{\it Phys.\ Rep.}\ {\bf #1}\ (#2) #3}
\def\PRL #1 #2 #3 {{\it Phys.\ Rev.\ Lett.}\ {\bf #1}\ (#2) #3}
\def\RMP #1 #2 #3 {{\it Rev.\ Mod.\ Phys.}\ {\bf #1}\ (#2) #3}
\def\ZP #1 #2 #3 {{\it Z.\ Phys.}\ {\bf #1}\ (#2) #3}

\begin{document}
\begin{flushright}
JINR E2-96-431\\
hep-ph/9612301
\end{flushright}
\vspace*{1cm}

\begin{center}
{\large \bf Double-Spin Asymmetry of  $J/\psi$ Production in Polarized pp \\
Collisions at HERA-$\vec{\rm N}$}\\
\vspace*{5mm}
{O.~TERYAEV\footnote{E-mail: teryaev@thsun1.jinr.dubna.su},
  A.~TKABLADZE\footnote{E-mail: avto@thsun1.jinr.dubna.su}}\\
{\it Bogoliubov Laboratory of Theoretical Physics,}\\
{\it JINR, Dubna, Moscow Region, 141980, Russia}\\
\end{center}

\begin{abstract}
We calculated the color-octet contribution to the double spin asymmetry
of $J/\psi$ hadroproduction with nonzero transverse momenta at
fixed target energies $\sqrt{s}\simeq40$ GeV.
It is shown that the color-octet contribution is dominant in the asymmetries.
The expected asymmetries and statistical errors
in a future option of HERA with longitudinally
polarized protons at HERA-$\vec N$ should allow one to
distinguish between different parametrizations for the polarized gluon
 distribution in the proton.
\end{abstract}

\newpage

\section{Introduction}
\setcounter{equation}{0}

The presently most accurate way to measure the polarized gluon distribution
function in the nucleon is to study those processes which can be calculated
in the framework of perturbative QCD (PQCD), i.e. for which the involved
production cross sections and subprocess asymmetries can be predicted.
One of the cleanest ways to probe QCD is to investigate heavy quarkonia
production processes.  Heavy quark pair production processes can be
 controlled perturbatively due to large mass of constituents.
On the other hand, heavy quark systems are mainly produced 
 in gluon fusion processes and therefore, asymmetries are expected to
 be sensitive to the polarized gluon  distribution in the nucleon.
 Investigation of heavy quarkonia production
 processes in polarized experiments would also yield additional
 information about the quark-antiquark pair hadronization phase.

 The two-spin asymmetry in $J/\psi$ production has been studied in the
 framework of the so called color singlet model (CSM) \cite{CSM} by Morii and
  collaborators \cite{MTY}. But as was shown in the last years, the color singlet
  model does not describe satisfactorily the heavy quarkonium hadroproduction
  at Tevatron and also at fixed target energies.
  While the CSM  gives a reasonable
description of the $J/\psi$ production cross section distribution shapes over
 $p_T$ or $x_F$ at fixed target energies, it completely fails
  in the explanation of the integrated cross
 section (a K factor 7-10 is needed to explain experimental data) \cite{FNAL}.
The anomalously large cross section \cite{CDF} of the $J/\psi$
 production at large transverse
momenta at the Tevatron revealed another negative feature of the CSM.
Within the framework of the CSM it is impossible to explain the anomalously
large $\psi$ \cite{BF} and direct $J/\psi$ production \cite{CGMP}
in the CDF experiment at the Tevatron.

The CSM is a nonrelativistic model where the relative velocity between
the heavy constituents in the bound state is  neglected. But discrepancies
between experimental data and the CSM predictions hint that $O(v)$
corrections as
well as other mechanisms  of quarkonium production, which do not appear
in the leading order in $v$, should be considered.
  Expansion of quarkonium cross sections and decay widths
in  powers of relative velocity $v$ of  heavy quarks in a bound state
has recently been realized in terms of Nonrelativistic
QCD (NRQCD) \cite{NRQCD}.
This formalism implies not only color-singlet processes but the new
color-octet mechanism, when a quark-antiquark pair is produced on  small
time scales in  color octet states and evolves into a  hadron by emission of
soft gluons.  The color octet mechanism takes into account the
 complete structure of the quarkonium Fock space while in the CSM 
 only the dominant Fock state is considered, which consists of a color singlet
 quark antiquark pair in the definite angular-momentum state
 (higher order Fock states are suppressed by powers of $v$).
According to the factorization approach based on the NRQCD,
the production cross section for a quarkonium state  H in the process
\begin{eqnarray}
 A+B\to H+X
\end{eqnarray}
can be written as
\begin{eqnarray}
  \sigma_{ij} = \sum_{i,j}{\int_{0}^{1} {dx_1 dx_2 f_{i/A}(x_1) f_{j/B}(x_2)
\hat\sigma(ij\to H)}}\\
\hat\sigma(ij\to H) = \sum_{n}{C^{ij}[n]\langle0|{\cal O}^{H}[n]|0\rangle}\nonumber
\end{eqnarray}
where  $f_{i/A}$ is the distribution function of the
parton $i$ in the hadron $A$.
The subprocess  cross section is separated into two parts: short 
distance~ ($C^{ij}[n]$) coefficients
and~ long~ distance matrix elements $\langle0|{\cal O}^H[n]|0\rangle$. The $C^{ij}[n]$ is the production
cross section of a heavy quark-antiquark pair in the $i$ and $j$
 parton fusion.
It should be calculated in the framework of pQCD. The [n] state can be
either  a color
singlet or a color octet state. The $\langle0|{\cal O}^H[n]|0\rangle$ describes the evolution
of a quark-antiquark pair into a hadronic state. These matrix elements cannot
be computed perturbatively. But the relative importance of long distance matrix
elements in powers of velocity $v$ can be estimated by using
the NRQCD velocity scaling rules \cite{LMNMH}.

Shapes of the $p_T$ distribution of short distance color-octet matrix elements
indicate that the new mechanism can explain the Tevatron data of direct
$J/\psi$ and $\psi'$ production at large $p_T$ \cite{CL}.
But unlike color-singlet matrix elements connected with the subsequent
hadronic nonrelativistic wave functions at the origin, color octet long
distance matrix elements are unknown and should be extracted from
experimental data.
The color octet contribution to the $J/\psi$ photoproduction has been analyzed
in the papers \cite{CK,AFM}. Recently, the $J/\psi$ hadroproduction at fixed
target energies has been studied by including the color-octet mechanism
\cite{GuS,BeR}. Large discrepancies between experimental data and the CSM
predictions for the total cross section of the
$J/\psi$ hadroproduction were explained.
The color octet contribution is dominant in the $J/\psi$ hadroproduction at
energies $\sqrt{s}\simeq30-60$ GeV.
The analyses carried out in these papers \cite{CK,AFM,GuS,BeR}
demonstrate that  fitting the photoproduction and hadroproduction data at
low energies requires  smaller values than those extracted from
 prompt $J/\psi$ production at CDF at the Tevatron \cite{CL}.
Possible reasons for such discrepancies have  recently been analyzed in the
papers \cite{BeR,ST}. The extraction of color-octet long distance matrix
 elements from the quarkonium production properties in polarized
 experiments would be an additional test of universality.

In the present letter we consider the color octet contribution to the double
spin asymmetry of
$J/\psi$ hadroproduction. Unlike the previous calculations \cite{GM}
we consider $(c\bar c)$ color octet and color singlet pair
  production in $2\to2$ subprocesses to obtain  asymmetries at nonzero
  transverse momenta ($p_T>1.5$ GeV). Such values of $p_T$ can not be
  caused by internal motion of partons in the nucleon and hence 
transverse momentum
  distributions of production cross section and asymmetries are calculable
  perturbatively. The double spin asymmetries in parton collisions are
  presented in section 2.
 Since heavy quarkonium is mainly produced in the gluon-gluon fusion
 subprocesses, the  $J/\psi$ production
asymmetry should be sensitive to the polarized gluon distribution
function in the proton.
We have calculated the expected asymmetry of $J/\psi$ production
 at HERA-$\vec N$, one of the future options of HERA \cite{Nowak};
an experiment utilizing an internal polarized nucleon target in the  polarized
HERA beam with energy $820$ GeV would yield $\sqrt{s}\simeq39$ GeV.
For comparison, we also considered the expected  asymmetry of $J/\psi$
 production in  similar spin physics experiments at much higher
  energies at the RHIC collider \cite{RHIC}.

\section{Double Spin Asymmetries at Subprocess Level}

Let us discuss the two-spin asymmetry $A_{LL}$ for the inclusive $J/\psi$ production
which is defined as
\begin{eqnarray}
A_{LL}^{J/\psi}(pp) = \frac{
d\sigma(p_+p_+\to J/\psi)-d\sigma(p_+p_-\to J/\psi)}
{d\sigma(p_+p_+\to J/\psi)+d\sigma(p_+p_-\to J/\psi)}=
\frac{Ed\Delta\sigma/d^3p}{Ed\sigma/d^3p}.
\end{eqnarray}
where $p_+(p_-)$ denotes helicity projection sign on the proton momentum
direction.

 The cross section of $J/\psi$ production can be written as
\begin{eqnarray}
\sigma_{J/\psi} = \sigma(J/\psi)_{dir} +
\sum_{J=0,1,2}{Br(\chi_{cJ}\to J/\psi X)\sigma_{\chi_{cJ}}}+Br(\psi'\to J/\psi X)
\sigma_{\psi},
\end{eqnarray}
where $Br((c\bar c)\to J/\psi X)$ denotes the branching ratio of the
corresponding $(c\bar c)$ state into $J/\psi$.
The production of each state of quarkonium is contributed by both color
octet and color singlet states, as in the case of direct $J/\psi$
production
\begin{eqnarray}
\sigma(J/\psi)_{dir} = \sigma_{J/\psi}^{singl}+\sigma^8_{J/\psi}=
\sigma(J/\psi)^{singl}+
\sum{\sigma(Q\bar Q[^{2s+1}L_J^{8}])
\langle0|{\cal O}_8^{J/\psi}(^{2s+1}L_J)|0\rangle}
\end{eqnarray}
where the sum is over the states $^3P_{0,1,2}^8$, $^1S_0^8$ and $^3S_1^8$.
We consider only the dominant sets of color octet states by the NRQCD
velocity expansion for the direct $S$ and $P$ state charmonium production.

In a recent paper \cite{GM} was considered asymmetry of $J/\psi$
hadroproduction in the color octet model exploiting only the lowest order
subprocesses ($2\to(c\bar c)$) over the QCD coupling constant.
Subprocesses $2\to1$ contribute
only to the production of a quarkonium state at zero transverse momentum with
respect to the beam axis. Transverse momentum distributions of $(c\bar c)$
states and, consequently, the $J/\psi$ meson are not calculable in the 
$p_T<\Lambda_{QCD}$ region. To deal with the experimentally observable
quantities (taking into account the kinematic restriction
on the angle with respect to the beam axis at HERA-$\vec N$ and at collider
experiments),
we consider the subprocess $2\to2$ which gives the leading contribution
to the quarkonium production with $p_T$ greater than $1.5$ GeV.
Such  large transverse momenta cannot be caused by internal motion of
partons in the nucleon and, respectively, subprocesses $2\to1$ should not
contribute to the production of quarkonia with $p_T>1.5$ GeV.
For calculating the expected asymmetries we consider only the following
subprocesses:
\begin{eqnarray}
 g+g &\to& (c\bar c)+ g\nonumber \\
 g+q &\to& (c\bar c)+ q
\end{eqnarray}
for the color octet and singlet states of a $(c\bar c)$-pair.
The quark-antiquark collision subprocesses are not taken into account,
because of small sea quark polarization in the proton.

 Using the symbolic manipulation program FORM \cite{FORM},
we have calculated the   $(c\bar c)$-states
production cross sections for different helicity states of colliding
partons. 
For the color octet states total cross sections have been calculated
by Cho and Leibovich \cite{CL}.
 The total cross sections for color singlet states are presented
in  \cite{BR,GTW}. The latter results served for checking our
calculations.
\begin{figure}[t]
\setlength{\unitlength}{1cm}
\begin{picture}(14,14)
\put(0,0){\epsfig{file=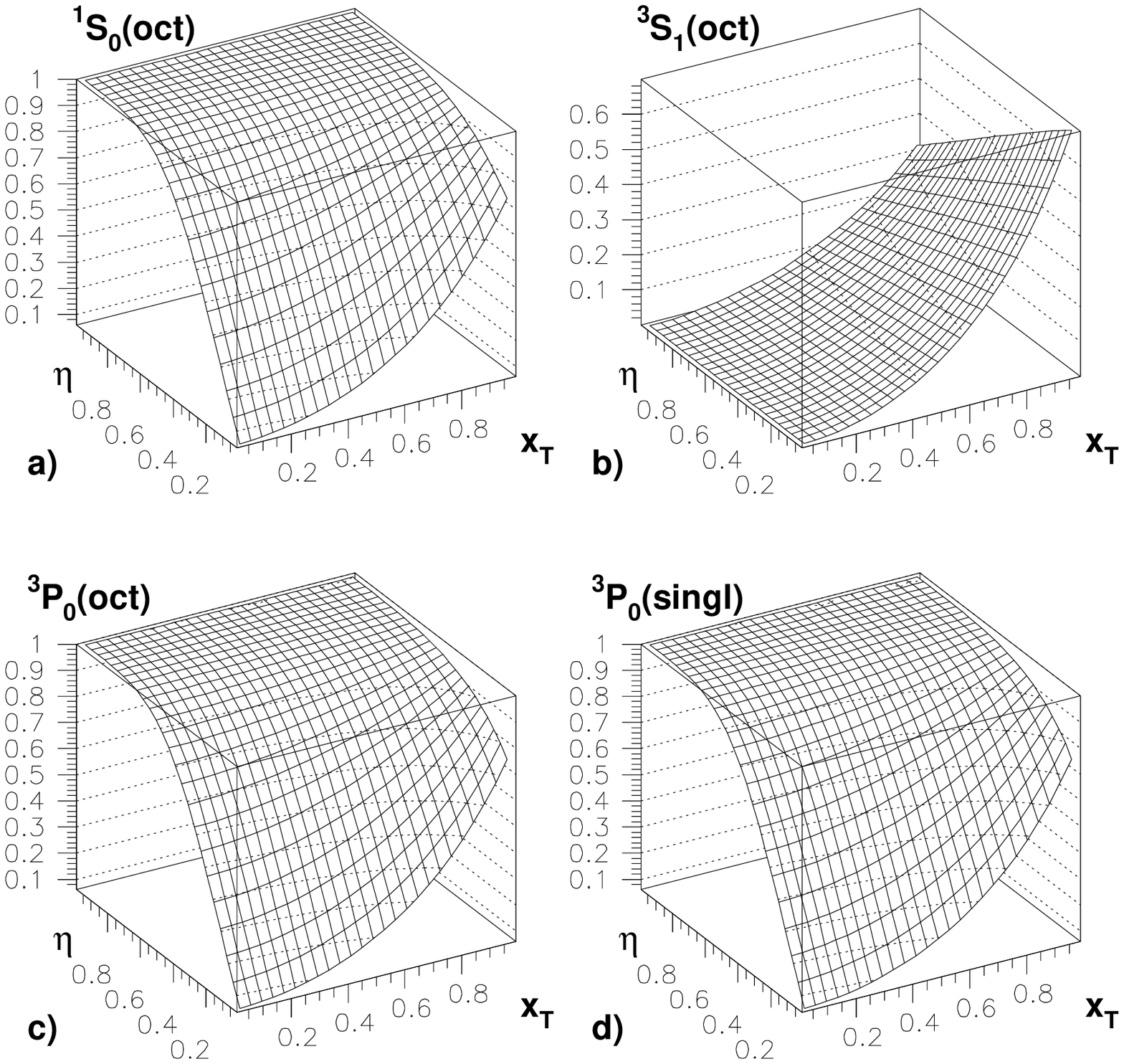,width=6.5in}}
\end{picture}

\noindent{\small
Figure~1. Partonic level double spin asymmetries for $^1S_0$ and $^3S_1$
color octet and $^3P_0$ color octet and singlet states versus $\eta$
and $x_T$.
}
\end{figure}

Figs.1 and 2 present the values of the subprocess level asymmetries
for the color octet and color singlet states
production of a ($c\bar c$)-pair
depending on  the two dimentionless
quantities: $\eta=4 m_c^2/\hat s^2$ and $x_T=p_T/p_{max}$, where $p_{max}$
is the maximum momentum of the produced state in the subprocess.
In  Figs.1 and 2  are shown only the gluon fusion subprocess asymmetries
because they give the main contribution to the hadronic level asymmetry.
The cases of $^1S_0$ and $^3S_1$ singlet states are omitted in  figs.1 and 2.
The $^1S_0$ state does not contribute to $J/\psi$ production and analytical
expressions for the corresponding cross sections for the $^3S_1$ state
are given in \cite{MTY}.

\begin{figure}[t]
\setlength{\unitlength}{1cm}
\begin{picture}(14,14)
\put(0,0){\epsfig{file=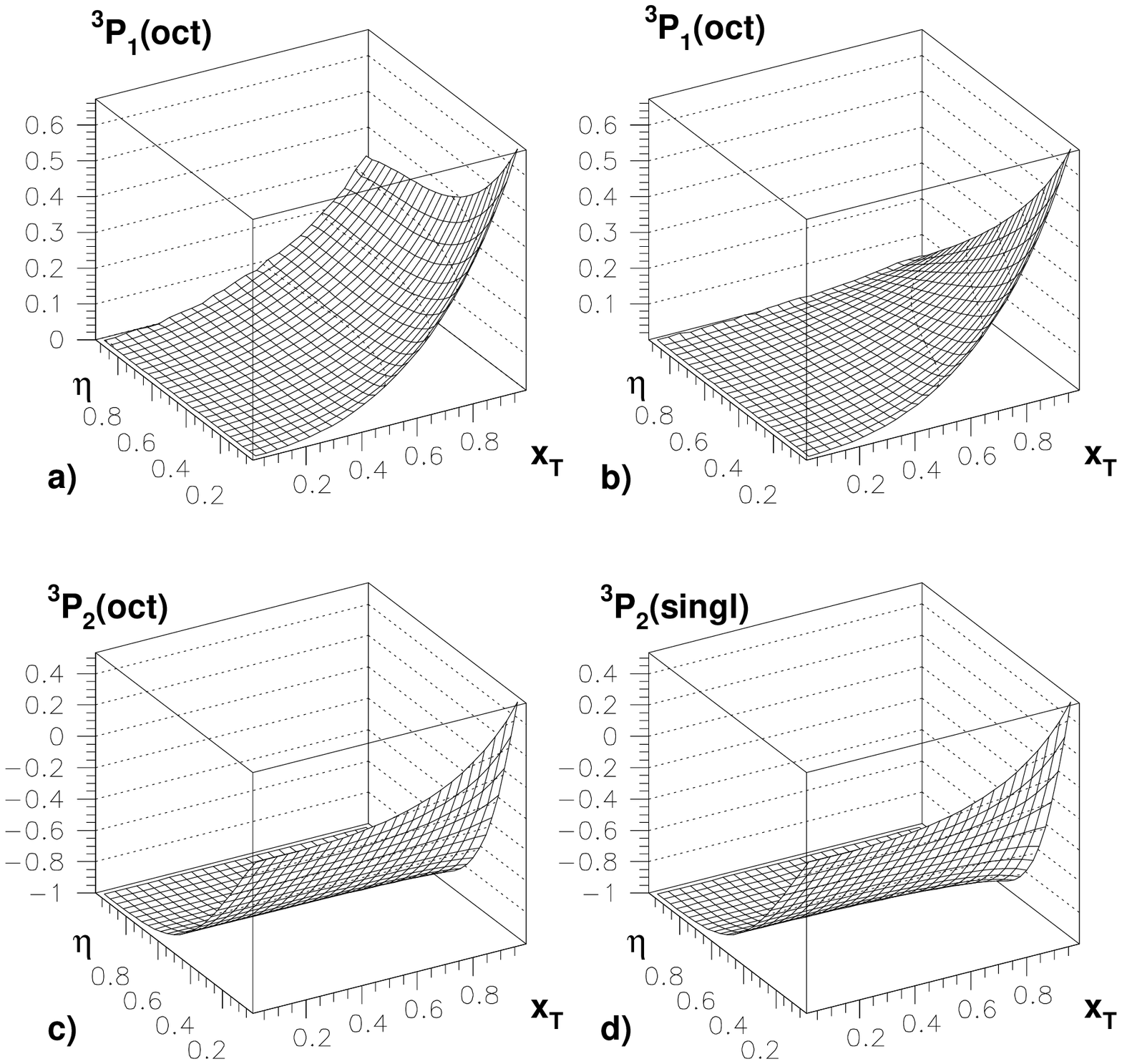,width=6.5in}}
\end{picture}

\noindent{\small
Figure~2. Partonic level double spin asymmetries for $^3P_1$ and $^3P_2$
color octet and singlet states versus $\eta$
and $x_T$.
}
\end{figure}

As can be seen from figs.1 and 2, the asymmetries are very similar for color octet and
color singlet states with the same spin-orbital quantum numbers.
It is worth  mentioning that in the limit $4m_c^2/s\to1$ (i.e. at the
threshold of heavy quark pair production) $\hat a_{LL}$ for the scalar
and tensor states tends to 1 and -1, respectively. The limit $4 m^2_c\to s$
means that the emitted gluon in the $2\to2$ subprocess ($gg\to(c\bar c)g$)
becomes soft and the helicity properties of the amplitude should be the same
as those of the amplitude of the $2\to1$ process ($gg\to(c\bar
c$) \footnote{The same reason makes the asymmetry less sensitive
 to the radiative corrections, than spin-averaged cross section, as K-factor
 is dominated by the contributions of virtual, soft and collinear gluons}. 
It is easy
to show that the asymmetries for  the process $2\to1$ 
 for scalar and tensor states
are 1 and -1, respectively. The existence of such limits serves as
an additional test for
our analytical calculations of cross sections $\Delta\hat\sigma$.
for scalar and tensor states
are 1 and -1, respectively. The existence of such limits serves as
an additional test for
our analytical calculations of cross sections $\Delta\hat\sigma$.

\section{Matrix Elements}

For the calculation of the hadronic level asymmetries of
$J/\psi$ production
we used the long distance
matrix elements fitted from various experimental data.
All  color singlet matrix elements are related to the radial quarkonium wave
functions at the origin and their derivatives. As in  paper \cite{CL}
for this purpose we used the Buchm\"uller-Tye wave functions values
at the origin tabulated in ref.\cite{BeR}.
The color octet matrix elements were fitted from the $J/\psi$  and higher
charmonium state production data in  various experiments.
Unfortunately, there are some discrepancies between the values of color octet
matrix elements extracted from different experiments.

 The number of color octet long distance matrix elements should be reduced by
 using the NRQCD spin symmetry relations:
\begin{eqnarray}
 \langle0|{\cal O}_8^H(^3P_J)|0\rangle &=& (2J+1)\langle0|{\cal O}_8^H(^3P_0)|0\rangle,\\
 \langle0|{\cal O}_8^{\chi_{cJ}}(^3S_1)|0\rangle &=& (2J+1)\langle0|{\cal O}_8^{\chi_{c0}}(^3S_1)|0\rangle.
\end{eqnarray}
These relations are accurate up to $v^2$.

After~ using~ these~ relations~ we have only four independent matrix elements
$\langle{\cal O}_8^{J/\psi}(^3S_1)\rangle$, $\langle{\cal O}_8^{\chi_{c1}}(^3S_1)\rangle$,
$\langle{\cal O}_8^{J/\psi}(^3P_0)\rangle$ and $\langle{\cal O}_8^{J/\psi}(^1S_0)\rangle$, which give the main
contributions to the $J/\psi$ hadroproduction cross section.
For the first two parameters we used the values extracted from Tevatron
data by Cho and Leibovich \cite{CL}.  The other parameters, connected with the
$\psi'$ production, were also taken from \cite{CL}.
As concerns the parameters
$\langle{\cal O}^{J/\psi}_8(^1S_0)\rangle$ and
$\langle{\cal O}^{J/\psi}_8(^3P_0)\rangle$, it is possible to extract only their
combination from direct $J/\psi$ production data at Tevatron \cite{CL}:
\begin{eqnarray}
\langle0|{\cal O}_8^{J/\psi}(^1S_0)|0\rangle+\frac{3}{m_c^2}\langle0|{\cal O}_8^{J/\psi}(^3P_0)|0\rangle =
6.6\cdot10^{-2} GeV^3.
\label{high}
\end{eqnarray}
Values for  the other combinations are extracted from the $J/\psi$ photoproduction
and fixed target hadroproduction data \cite{CK,AFM,GuS,BeR}
\begin{eqnarray}
\langle0|{\cal O}_8^{J/\psi}(^1S_0)|0\rangle+\frac{7}{m_c^2}\langle0|{\cal O}_8^{J/\psi}(^3P_0)|0\rangle =
2\cdot10^{-2} GeV^3 \cite{BeR},\nonumber\\
\langle0|{\cal O}_8^{J/\psi}(^1S_0)|0\rangle+\frac{7}{m_c^2}\langle0|{\cal O}_8^{J/\psi}(^3P_0)|0\rangle =
3\cdot10^{-2} GeV^3 \cite{CK,AFM}.
\label{low}
\end{eqnarray}
If one assumes that
$\langle0|{\cal O}_8^{J/\psi}(^1S_0)|0\rangle=\langle0|{\cal O}_8^{J/\psi}(^3P_0)|0\rangle/m_c^2$
the photoproduction and fixed target hadroproduction values ('low energy'
values) are  an order  smaller  than the  Tevatron value ('high energy' value).
The possible sources for such a discrepancy are discussed in
\cite{BeR,ST}. In  paper \cite{BeR} it is mentioned that the mass of 
the produced hadronic final 
state (or intermediate color octet state) must be higher
than that of the $J/\psi$ meson -- $M_{J/\psi}=2m_c$, because the intermediate
octet state emits  gluons with energy $2m_c v^2$
before transition into a color singlet state.
But a charmonium is not a truly nonrelativistic system, the average
relative velocity of constituents in  $J/\psi$ is not very small ---
$v^2\simeq0.23-0.3$ and  emitted gluons
in the $c\bar c$ pair hadronization phase have the energy  $0.7\div$ GeV.
Therefore the mass of the  hadronic final state is about $4$ GeV.
At fixed target energies, increasing  the mass of the produced hadronic state
leads to a  reduced  partonic luminosity because of a steeply falling
gluon distribution function. Consequently, the production cross section
of the ($c\bar c$) color octet state should  be smaller. So, the 'true'
color octet long distance matrix elements must be larger, than those
extracted by using  $M_{J/\psi}$ as mass of the intermediate octet state
\cite{BeR}.

  Another possible source was mentioned in the paper \cite{ST} and
should be an uncertainty connected with the choice
of different parametrizations for the gluon distribution function in
fitting the direct $J/\psi$ production data at CDF \cite{CL}.
Cho and Leibovich used in their calculations  the MRSD0
parametrization for parton distribution functions.
At small values of $p_T\simeq5$ GeV, using  a more reliable parametrization
for small partonic $x$ (GRV LO or  GRV HO\cite{GRV}) leads to 
$1.5\div1.6$ times higher  cross sections
for  the $^3P_0$ and $^1S_0$ color octet state production than
those obtained by using MRSD0 parametrization \cite{CL}.
Hence the fitted value of
  combination (\ref{high}) should be approximately $1.5\div1.6$ times smaller:
\begin{eqnarray}
\langle0|{\cal O}_8^{J/\psi}(^1S_0)|0\rangle+\frac{3}{m_c^2}\langle0|{\cal O}_8^{J/\psi}(^3P_0)|0\rangle =
4\div4.4\cdot10^{-2} GeV^3 \cite{ST}.
\label{MY}
\end{eqnarray}
At large $p_T\simeq10\div15$ GeV all parametrizations give practically
the same magnitude for the color octet states cross sections.
Hence if we use the new combination (\ref{MY})
one needs a larger value of the parameter $\langle0|{\cal O}^{J/\psi}_8(^3S_1)|0\rangle$
to explain the  experimental data for the $J/\psi$ cross section at
$p_T\simeq10\div15$ GeV:
\begin{eqnarray}
   \langle0|{\cal O}^{J/\psi}_8(^3S_1)|0\rangle \simeq 10\cdot10^{-3} GeV^3.
\end{eqnarray}
Practically the same values for these parameters  were obtained
by Beneke and Kr\"amer in a  recent paper by fitting the CDF data of direct
$J/\psi$ production and using the GRV LO (1994) parametrization \cite{BK}:
\begin{eqnarray}
 &&  \langle0|{\cal O}^{J/\psi}_8(^3S_1)|0\rangle= 1.06\pm0.14^{+1.05}_{-0.59}\cdot10^{-2}
    GeV^3,\\
&&\langle0|{\cal O}_8^{J/\psi}(^1S_0)|0\rangle+\frac{3.5}{m_c^2}\langle0|{\cal O}_8^{J/\psi}(^3P_0)|0\rangle =
3.9\pm1.15^{+1.46}_{-1.07}\cdot10^{-2} GeV^3.
\end{eqnarray}

For  calculations of the hadronic level asymmetries we use three different
sets for the three long-distance color octet matrix elements

\begin{eqnarray}
&A)& -~\langle0|{\cal O}_8^{J/\psi}(^3S_1)|0\rangle=6.6\cdot10^{-3} GeV^-3 \cite{CL},\nonumber\\
&&\langle0|{\cal O}_8^{J/\psi}(^1S_0)|0\rangle+\frac{3}{m_c^2}\langle0|{\cal O}_8^{J/\psi}(^3P_0)|0\rangle =
6.6\cdot10^{-2} GeV^3~~ \cite{CL};\nonumber\\
&~&\nonumber\\
&B)& -~\langle0|{\cal O}_8^{J/\psi}(^3S_1)|0\rangle=6.6\cdot10^{-3} GeV^-3 \cite{CL},\nonumber\\
&&\langle0|{\cal O}_8^{J/\psi}(^1S_0)|0\rangle+\frac{3}{m_c^2}\langle0|{\cal O}_8^{J/\psi}(^3P_0)|0\rangle =
3\cdot10^{-2} GeV^3~~ \cite{BeR};\nonumber \\
&~&\nonumber\\
&C)& -~\langle0|{\cal O}_8^{J/\psi}(^3S_1)|0\rangle=10\cdot10^{-3} GeV^-3 \cite{ST},\nonumber\\
&&\langle0|{\cal O}_8^{J/\psi}(^1S_0)|0\rangle+\frac{3}{m_c^2}\langle0|{\cal O}_8^{J/\psi}(^3P_0)|0\rangle =
4\cdot10^{-2} GeV^3~~ (\ref{MY}).\nonumber
\end{eqnarray}
As has been mentioned above, values for the other color octet parameters were taken
from \cite{CL}.  For  the calculations of the expected asymmetries we
assume that the leading term in the combination (\ref{MY}) is the
parameter
$\langle{\cal O}^{J/\psi}_8(^1S_0)\rangle$ (or
$\langle{\cal O}^{J/\psi}_8(^3P_0)\rangle=0$). Only wiyh  such a radical choice the 
values of 'high' and 'low' energy  parameters should be consistent
to each other.

\section{Results and Discussion}
Fig. 3a presents the expected asymmetries for the three sets of color octet
parameters in the case  when the parameter
$\langle{\cal O}^{J/\psi}_8(^3P_0)\rangle$ tends to zero ($\langle{\cal O}^{J/\psi}_8(^1S_0)\rangle$ is the
leading term in the combinations (\ref{high}),(\ref{low}),(\ref{MY})).
\begin{figure}[h]
\setlength{\unitlength}{1cm}
\begin{picture}(16,7.5)
\put(0,0){\epsfig{file=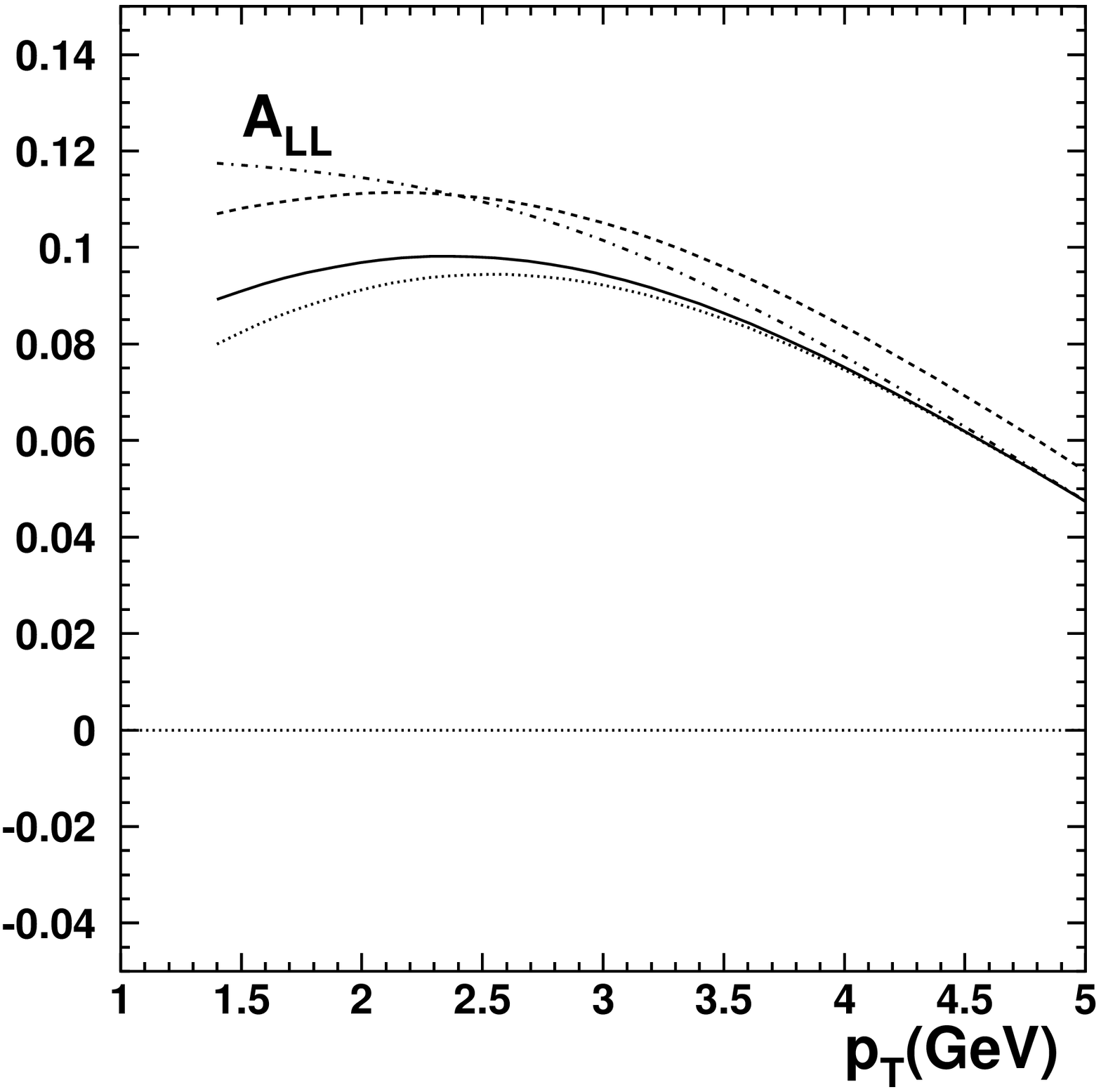,width=7.5cm}}
\put(8,0){\epsfig{file=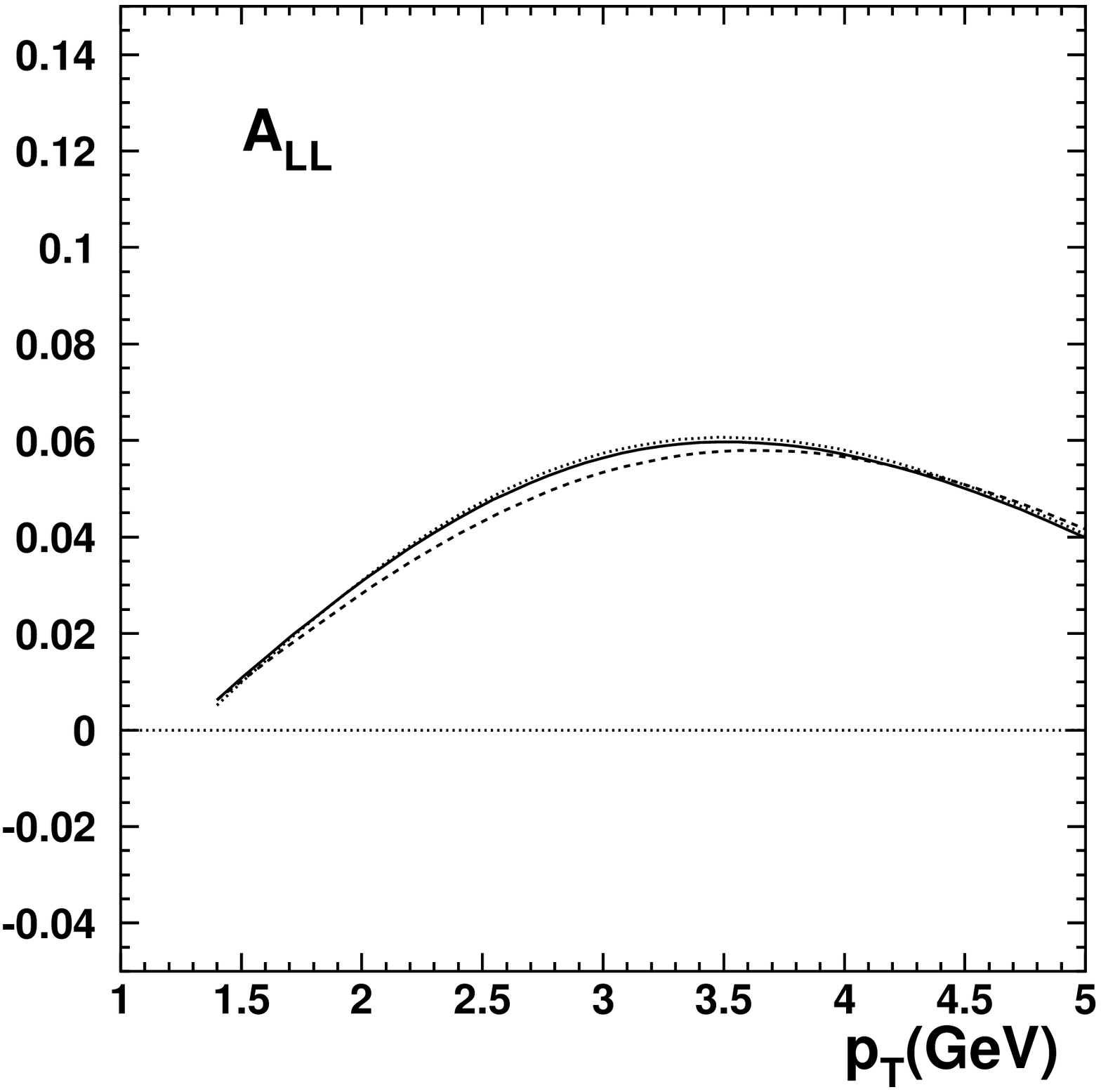,width=7.5cm}}
\end{picture}

\noindent{\small
Figure~3. The double spin asymmetries for different sets of
color-octet  long distance parameters. Solid line corresponds to set $C$,
dashed line -- set A, dotted line -- set B; (a) for the case
when  $\langle0^{J/\psi}_8(^3P_0)\rangle=0$, (b)for --  $\langle0^{J/\psi}_8(^1S_0)\rangle=0$.
Dot-dashed line in Fig.a corresponds to asymmetry calculated by using
values of parameters from \cite{FleN}.
}
\end{figure}
Fig. 3b represents the other radical choice, when the  second parameter
$\langle{\cal O}^{J/\psi}_8(^1S_0)\rangle$ is zero.
For the calculations of hadronic level asymmetries we used the GRV LO
parametrization for unpolarized distribution functions \cite{GRV} and
a parametrization proposed by Gehrmann and Stirling for the polarized parton
distribution function (set A) \cite{GS}. For the mass of the charm quark
the value $m_c=1.48$ GeV was taken.
As we can see from Fig.3, the magnitude of the asymmetry is more
sensitive to the choice of the leading term in the combinations than
to the choice of the particular set of the octet matrix elements.
 Recently the color octet matrix elements
$\langle{\cal O}^{J/\psi}_8(^3P_0)\rangle$ and $\langle{\cal O}^{J/\psi}_8(^1S_0)\rangle$ have been
extracted separately from the $J/\psi$ electroproduction data \cite{FleN}.
\begin{eqnarray}
\langle0|{\cal O}^{J/\psi}_8(^1S_0)|0\rangle & = & 4\cdot10^{-2} GeV^3, \nonumber\\
\frac{\langle0|{\cal O}^{J/\psi}_8(^3p_0)|0\rangle}{m_c^2}
 & = & -0.3\cdot10^{-2} GeV^3.
\label{Eprod}
\end{eqnarray}
These values confirm our assumption that the leading term in the combinations
is the parameter $\langle{\cal O}^{J/\psi}_8(^1S_0)\rangle$.  In  fig.3a, the  dotted-dashed
curve corresponds to the asymmetry calculated by using the
values of the long-distance parameters (\ref{Eprod}). For the parameter
$\langle{\cal O}^{J/\psi}_8(^3S_1)\rangle$ a value from  set C was used.
As one can see from figs.3(a,b), the expected asymmetries for all sets of
parameters  are practically the same.
In the following calculations we shall use the set C for the values of
the parameters, assuming $\langle0|{\cal O}^{J/\psi}_8(^3P_0)|0\rangle=0$.
\begin{figure}[h]
\setlength{\unitlength}{1cm}
\begin{picture}(16,7.5)
\put(0,0){\epsfig{file=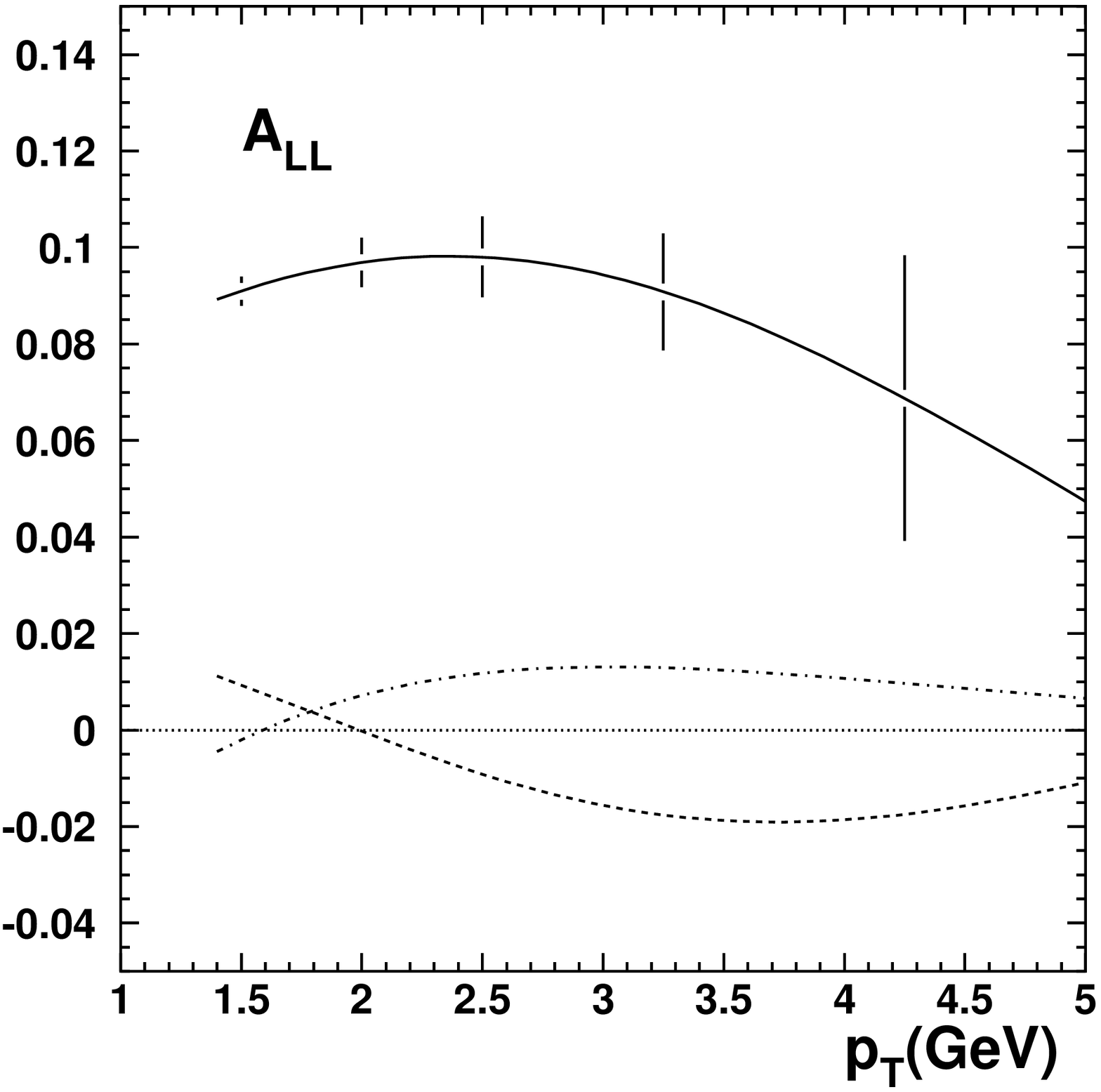,width=7.5cm}}
\put(8,0){\epsfig{file=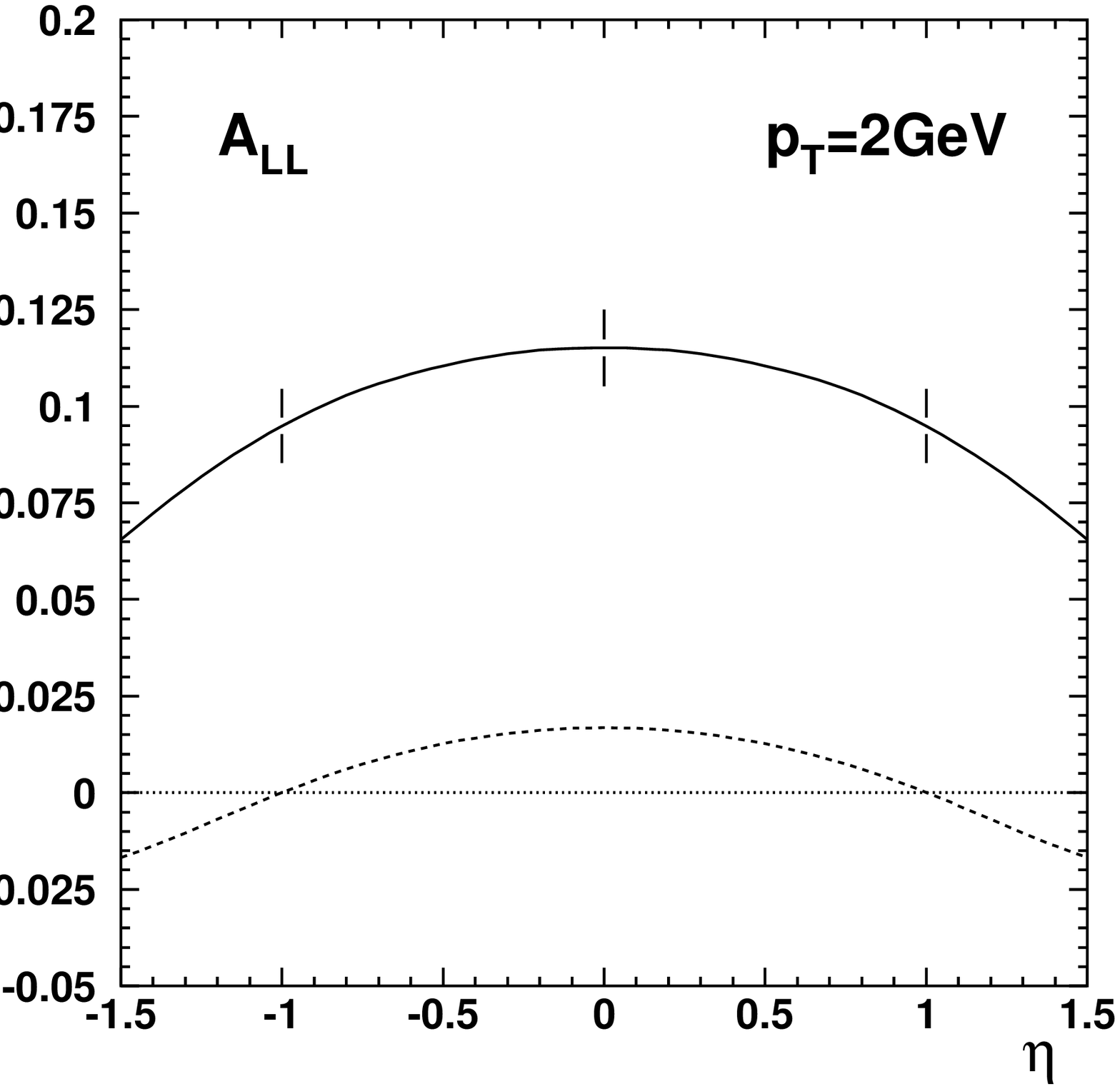,width=7.5cm}}
\end{picture}

\noindent{\small
Figure~4. The expected asymmetries at $\sqrt{s}=39$ GeV. Solid line
corresponds Gehrmann-Stirling  polarized parton parametrization,
set A; dashed line, set C \cite{GS}. (a) expected asymmetries versus
transverce momentum, (b) versus pseudorapidity of $J/\psi$. Dot-dashed
curve in the fig.a corresponds to the color-singlet contribution to asymmetry. 
}
\end{figure}
Fig.4a shows the expected double spin asymmetries at HERA-$\vec N$ energies
as functions of transverse momentum $J/\psi$ in the c.m.s.  The solid curve corresponds
to the Gehrman and Stirling parametrization for polarized distribution
functions, set A; the dashed curve, to set C \cite{GS}.
In  fig.4a we also display the expected statistical errors $\delta A_{LL}$
at HERA-$\vec N$ which can be estimated from \cite{Nowak}
\begin{eqnarray}
\delta A_{LL} = 0.17/\sqrt{\sigma(pb)}.
\end{eqnarray}
This relation has been determined by assuming an integrated luminosity of
$240~pb^{-1}$ and beam and target polarizations $P_B=0.6$,
$P_T=0.8$ \cite{Nowak}.
The error bars  are obtained by using integrated cross sections over bins
$\Delta p_T=0.5$ GeV (for the first three points) and $\Delta p_T=1$ GeV (for
the other two ones).
The $J/\psi$ decay branching ratio into the $e^+e^-$ mode is also included.
The magnitude of asymmetries and expected errors allows one to distinguish
between different parametrizations of polarized parton distribution functions.
Fig.4b shows expected asymmetries  depending on the
pseudorapidity of $J/\psi$ at $p_T=2$ GeV. Statistical error bars correspond
to the integrated cross section over bins $\Delta\eta=1$ and
$\Delta P_T=0.5$ GeV.  As in the previous case, the errors are small in a
wide range of pseudorapidity interval and give a possibility to
distinguish between different parametrizations for $\Delta g(x)$.
It is worth mentioning that gluon-quark collision subprocesses give
about $5\div10\%$ in  $\Delta\sigma$.
The dot-dashed curve in  fig.4a represents the color-singlet contribution
to the asymmetry for the parametrization set A (only the color singlet
contribution in $\Delta\sigma$ was taken into account). It is clear 
that the color octet
contribution dominates in the expected asymmetries of $J/\psi$
production. 
One may ask, where the such a difference in the shapes of asymmetries 
is coming from,
as the subprocess asymmetries are not sensitive to colour structure. 
The main reason for this is the large matrix element 
$<0|O^{J/\psi}_8(^1S_0)|0>$. The production of such a state 
has a large asymmetry $A \sim 1$ (at  
$\eta \sim 1$, which is not far from 
the reality for the considered $p_T$), providing the smooth 
transition to $2 \to 1$ process, discussed above. 
As such a state is absent in CSM, the large asymmetry itself
is already a sign of a presence of colour-octet contribution.

One of main parameters of the model is the mass of the charm quark.
Fig.5 shows the asymmetries  of $J/\psi$ production  dependening
on  $m_c$ for two values of $p_T$ for a large set of polarized parton
 parametrizations (set A).
As displayed in fig.5, the expected asymmetries are practically insensitive
to the  quark mass  above $m_c=1.5$ GeV.
\begin{figure}[h]
\setlength{\unitlength}{1cm}
\begin{picture}(16,7.5)
\put(3,0){\epsfig{file=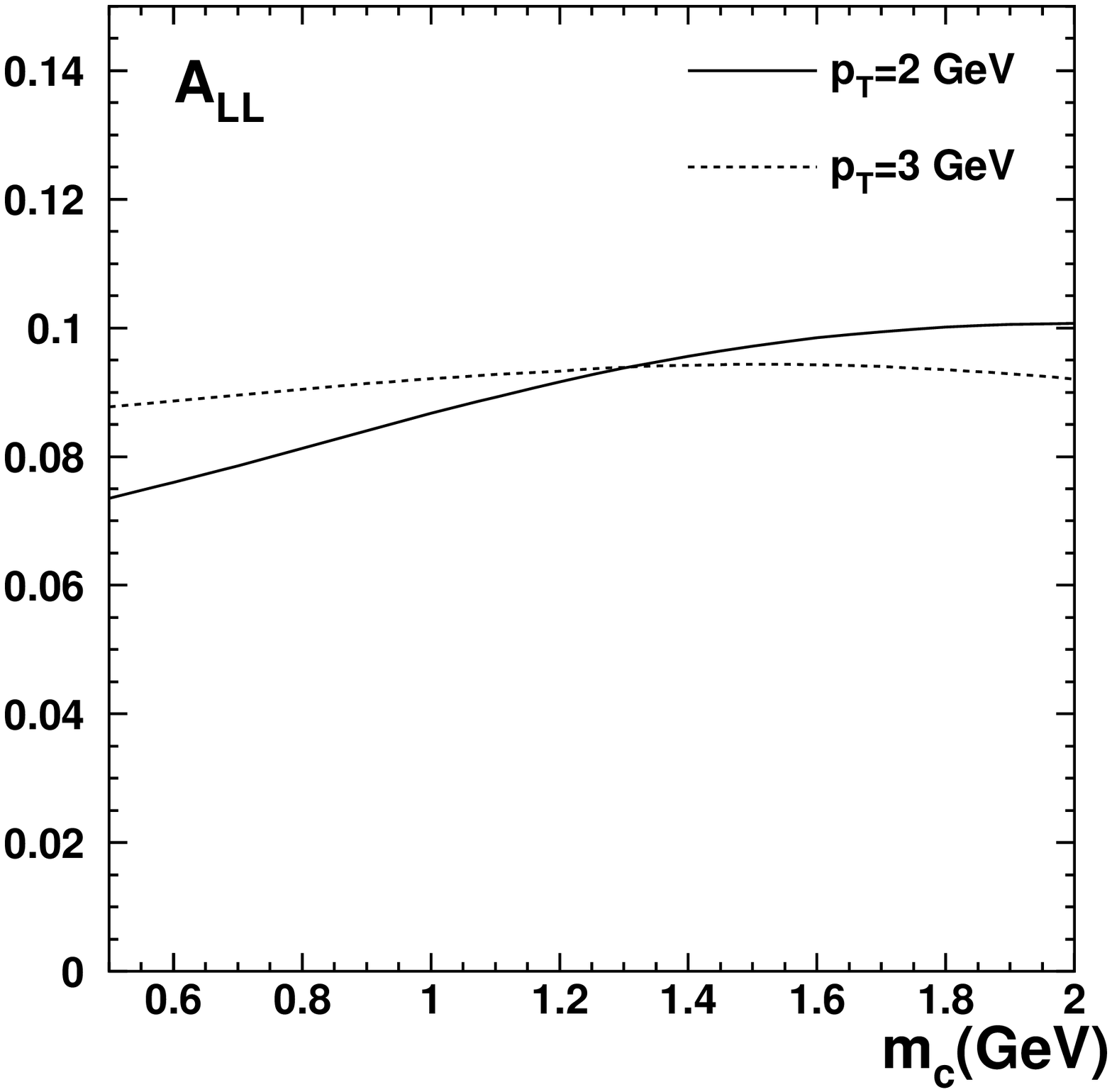,width=7.5cm}}
\end{picture}

\noindent{\small
Figure~5. Expected  double-spin asymmetries at  $\sqrt{s}=39$ GeV versus
mass of charm quark.
}
\end{figure}
Therefore, the double spin asymmetry of $J/\psi$ production,
unlike the cross section, should be free form uncertainties caused
by a unknown mass of the intermediate color octet states.

\begin{figure}[h]
\setlength{\unitlength}{1cm}
\begin{picture}(16,7.5)
\put(3,0){\epsfig{file=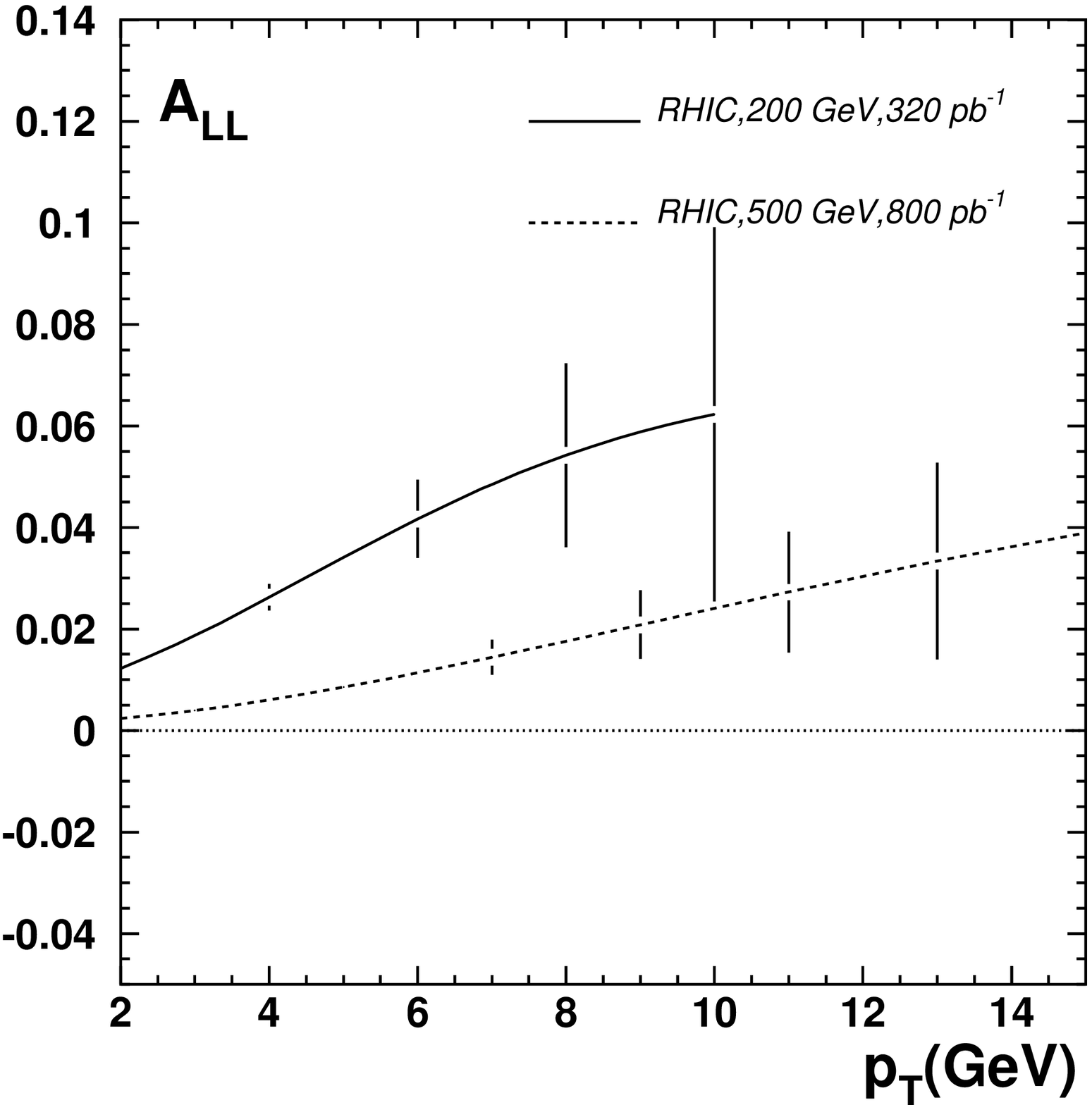,width=7.5cm}}
\end{picture}

\noindent{\small
Figure~6. Expected asymmetries and statistical errors at the RHIC for 
two different energies.
}
\end{figure}

For  comparison we have also calculated the expected double-spin
asymmetries of $J/\psi$   production at RHIC energies.
The results are given in Fig.6 for two different values of energy.
The expected statistical errors are calculated with the anticipated 
the integrated luminosities at the corresponding energies.
As we can see from Fig.6, the expected asymmetry decreases with 
increasing  c.m.s. energy. The statistical errors are calculated
by integration over $p_T$ with bins $\Delta p_T=0.5$ GeV
 of the differential cross sections.

\section{Conclusions}

In this paper we investigated the expected double spin asymmetries of
heavy quarkonium hadroproduction at HERA-$\vec N$.
To deal with experimentally
observed quantities, we
considered    $J/\psi$ meson production at nonzero transverse momenta,
$p_T>1.5$ GeV. Unlike the calculations of \cite{GM}, where only the lowest
order subprocesses were taken into account ($2\to1$), we
considered $J/\psi$ production in the subprocesses $2\to2$ because
large values of $p_T$ can not be caused by internal motion of partons.
 We have calculated
the  heavy quark pair color octet and color singlet $S$ and $P$ states
production cross sections for different helicities of colliding partons.
The FORTRAN codes for calculated cross sections are available by E-mail.
For  calculation of the expected hadronic asymmetries  we used  more
reliable values for the color-octet
long distance matrix elements, which are in good agreement with those
extracted from the $J/\psi$ electroproduction data \cite{FleN}.

 The magnitude of expected asymmetries
and  statistical errors at HERA-$\vec N$ allows one to distinguish
between different parametrizations for polarized parton distribution
functions (Gehrmann and Stirling, set A and C).  On the other hand,
measuring the asymmetry  would give a possibility to extract information
about the color-octet long distance matrix elements  and  to check universality
of factorization.
 We also calculated the $J/\psi$ production asymmetries at RHIC
 energies.
 By comparing the magnitudes of the expected asymmetries at HERA-$\vec N$ and
STAR, it becomes  clear that the energy of the fixed target experiment
 is more preferable
for the investigation of the charmonium production asymmetry.

{\bf Acknowledgements}

We are grateful to S.J.~Brodsky and W.-D.~Nowak for helpful discussions.
This work was supported in part by the Russian Foundation for
Fundamental Investigation under  grant $96-02-17631$.

\end{document}